\documentclass[aps,prl,twocolumn,showpacs,superscriptaddress]{revtex4-1}
\usepackage{graphicx}
\usepackage{amsmath,amsfonts}
\usepackage[breaklinks]{hyperref}

\newcommand{\exv}[1]{{\langle{#1}\rangle}}
\newcommand{\ket}[1]{{\lvert{#1}\rangle}}
\newcommand{\bracket}[2]{{\langle{#1}|{#2}\rangle}}
\newcommand{\proj}[1]{{\lvert{#1}\rangle\!\langle{#1}\rvert}}
\newcommand{\abs}[1]{{\lvert{#1}\rvert}}
\newcommand{\context}{{\underline{\mathfrak c}}}
\newcommand{\Context}{{\mathfrak C}}

\begin{document}

\title{Optimal inequalities for state-independent contextuality}

\author{Matthias Kleinmann}
 \email{matthias.kleinmann@uni-siegen.de}
 \affiliation{Naturwissenschaftlich-Technische Fakult\"at,
 Universit{\"a}t Siegen, Walter-Flex-Stra{\ss}e 3, D-57068 Siegen,Germany}

\author{Costantino Budroni}
 \email{cbudroni@us.es}
 \affiliation{Departamento de F\'{\i}sica Aplicada II,
 Universidad de Sevilla, E-41012 Sevilla, Spain}
 \affiliation{Naturwissenschaftlich-Technische Fakult\"at,
 Universit{\"a}t Siegen, Walter-Flex-Stra{\ss}e 3, D-57068 Siegen,Germany}

\author{Jan-{\AA}ke Larsson}
 \email{jan-ake.larsson@liu.se}
 \affiliation{Institutionen f\"or Systemteknik,
 Link\"opings Universitet, SE-58183 Link\"oping, Sweden}

\author{Otfried G{\"u}hne}
 \email{otfried.guehne@uni-siegen.de}
 \affiliation{Naturwissenschaftlich-Technische Fakult\"at,
 Universit{\"a}t Siegen, Walter-Flex-Stra{\ss}e 3, D-57068 Siegen,Germany}

\author{Ad\'an Cabello}
 \email{adan@us.es}
 \affiliation{Departamento de F\'{\i}sica Aplicada II,
 Universidad de Sevilla, E-41012 Sevilla, Spain}

\begin{abstract}
Contextuality is a natural generalization of nonlocality which does not need 
 composite systems or spacelike separation and offers a wider spectrum of 
 interesting phenomena.
Most notably, in quantum mechanics there exist scenarios where the contextual 
 behavior is independent of the quantum state.
We show that the quest for an optimal inequality separating quantum from 
 classical noncontextual correlations in an state-independent manner admits an 
 exact solution, as it can be formulated as a linear program.
We introduce the noncontextuality polytope as a generalization of the locality 
 polytope, and apply our method to identify two different tight optimal 
 inequalities for the most fundamental quantum scenario with state-independent 
 contextuality.
\end{abstract}


\maketitle

\emph{Introduction.}---%
The investigation of the operational differences between quantum mechanics and 
 classical mechanics resulted {1964} in the discovery of Bell's inequalities 
 \cite{Bell:1964PHY}.
Such inequalities constrain the correlations obtained from spacelike-separated 
 measurements and are satisfied by any local hidden variable (HV) model but are 
 violated by quantum mechanics.
For every measurement scenario, there exists a minimal set of inequalities, 
 called \emph{tight} Bell inequalities, which provide also sufficient 
 conditions:
If all tight inequalities are satisfied, then there exists a local HV model 
 reproducing the corresponding set of correlations \cite{Fine:1982PRL, 
 Pitowsky:1989}.

Mathematically speaking, each tight Bell inequality corresponds to a facet of 
 the locality polytope \cite{Pitowsky:1989}.
This means that it is an $(p-1)$-dimensional face of the $p$-dimensional 
 polytope obtained as a convex hull of the vectors representing local 
 assignments to the results of the considered measurements.
Such a polytope gives all classical probabilities associated to a local model 
 for a given measurement scenario, and its facets give precisely the boundaries 
 of the polytope.
In this sense, tight inequalities separate classical from nonclassical 
 correlations perfectly.

Similarly, noncontextuality inequalities \cite{Klyachko:2008PRL, 
 Cabello:2008PRL, Badziag:2009PRL} are constraints on the correlations among 
 the results of compatible observables, which are satisfied by any 
 noncontextual HV model.
While the violation of Bell inequalities reveals nonlocality, the violation of 
 noncontextuality inequalities reveals contextuality \cite{Specker:1960DIE, 
 Bell:1966RMP, *Kochen:1967JMM}, which is a natural generalization of 
 nonlocality privileging neither composite systems (among other physical 
 systems), nor spacelike-separated measurements (among other compatible 
 measurements), nor entangled states (among other quantum states).

All Bell inequalities are noncontextuality inequalities, but there are two 
 features of noncontextuality inequalities which are absent in Bell 
 inequalities.
One is that noncontextuality inequalities may be violated by simple quantum 
 systems such as single qutrits \cite{Klyachko:2008PRL}.
These violations have recently been experimentally observed with photons 
 \cite{Lapkiewicz:2011Nat, *AABC11}.
The other is that the violation can be independent of the quantum state of the 
 systems \cite{Cabello:2008PRL, Badziag:2009PRL}, thus it reveals 
 state-independent contextuality (SIC).
The latter has been demonstrated recently with ququarts (four-level quantum 
 systems) using ions \cite{Kirchmair:2009NAT}, photons \cite{Amselem:2009PRL}, 
 and nuclear magnetic resonance \cite{Moussa:2010PRL}.

The notion of tightness naturally applies also to noncontextuality 
 inequalities.
Tight noncontextuality inequalities are the facets of the correlation polytope 
 of compatible observables as we will explain below.
Compared with the locality polytope, the difference is in the notion of 
 compatibility, since now one no longer considers only collections of 
 spacelike-separated measurements, but admits more generally the measurement of 
 a \emph{context}, i.e., a collection of mutually compatible measurements.
For a given contextuality scenario, the corresponding set of tight inequalities 
 gives necessary and sufficient conditions for the existence of a noncontextual 
 model.

For example, the three inequalities with state-independent violation introduced 
 in Ref.~\cite{Cabello:2008PRL}, are all tight.
These inequalities are only violated for ququarts (two of the inequalities) and 
 eight-level quantum systems (the third inequality), but not for qutrits.
Another example of a tight inequality is the noncontextuality inequality for 
 qutrits of Klyachko {\em et al.} \cite{Klyachko:2008PRL}, which indeed was 
 derived by means of the correlation polytope method.
However, this latter inequality does not have a state-independent quantum 
 violation.

Obtaining all tight inequalities is, in general, a hard task.
The correlation polytope is characterized by the number of settings and 
 outcomes of the considered scenario.
While there are algorithms that find all the facets of a given polytope, the 
 time required to compute them grows exponentially as the number of settings 
 increases.
Therefore, this method can only be applied to simple cases with a reduced 
 number of settings \cite{Fine:1982PRL, Klyachko:2008PRL, Pitowsky:2001PRA, 
 *Collins:2004JPA}.
Given the facets of the polytope, in a next step one can try to find quantum 
 observables that exhibit a maximal gap between the maximal noncontextual value 
 and the quantum prediction.

In this paper we approach the problem differently.
For many situations, the quantum observables are already known, and it remains 
 to find inequalities that are tight and optimal and, in addition, may exhibit 
 SIC.
Thus we first describe the noncontextuality polytope for a given set of 
 observables and a given list of admissible contexts.
A noncontextuality inequality is then an affine hyperplane that does not 
 intersect this polytope.
We then introduce a method for maximizing the state-independent quantum 
 violation via linear programming.
The resulting linear program can be solved with standard optimization routines, 
 and the optimality of the solution is guaranteed.
As an application we derive the optimal inequality for several state 
 independent scenarios, in particular analyzing a recently discovered qutrit 
 scenario \cite{Yu:2012PRL}.
Using our method, we find noncontextuality inequalities with state-independent 
 violation and the fewest number of observables and contexts.
These inequalities turn out to be in addition tight and hence provide the most 
 fundamental examples of inequalities with state-independent violation.

\emph{Contextuality scenarios, the noncontextuality polytope, and 
noncontextuality inequalities.}---%
We start from some given dichotomic \footnote{%
Dichotomy is not a restriction, since using the spectral decomposition $X=\sum 
 \lambda_i \Pi_i$ the observable $X$ can be replaced by the dichotomic 
 observables $X_i= 2\Pi_i- \openone$.}
 quantum observables $A_1, A_2, \dotsc, A_n$.
A context $\context$ is then a set of indices, such that $A_k$ and $A_\ell$ are 
 compatible whenever $k, \ell\in \context$, i.e., $[A_k, A_\ell]= 0$.
For example if $A_1$ and $A_2$ are compatible, then valid contexts would be 
 $\{1\}$, $\{2\}$, and $\{1,2\}$.
As we see below, it may be interesting to consider only a certain admissible 
 subset $\Context$ of the set of all possible contexts $\{\context\}$.
The observables $A_1, \dotsc, A_n$, together with the list of admissible 
 contexts $\Context$, form the contextuality scenario.

The set of all (contextual as well as noncontextual) correlations for such a 
 scenario can be represented by the following standard construction.
We first use that, if $A_k$ and $A_\ell$ are compatible, then the expectation 
 value of $A_k$ is not changed whether or not $A_\ell$ is measured in the same 
 context.
Thus, instead of considering all correlations, it suffices to only consider the 
 vector $\vec v=(v_\context\mid \context\in \Context)$, where $v_\context$ is 
 the expectation value of the product of the values of the observables indexed 
 by $\context$.
For example, for the contexts $\{1\}$, $\{2\}$, $\{1,2\}$, a contextual HV 
 model may with equal probability assign the values $\{+1\}$, $\{+1\}$, 
 $\{-1,+1\}$, or $\{-1\}$, $\{-1\}$, $\{+1,-1\}$, respectively, yielding $\vec 
 v\equiv (v_1,v_2,v_{1,2})=(1/2,1/2,-1)$.

In the simplest \emph{noncontextual} HV model, however, each observable has a 
 fixed assignment $\vec a\equiv (a_1, \dotsc, a_n)\in \{-1,1\}^n$ for the 
 observables $A_1, \dots, A_n$, and accordingly each entry in $\vec v$ is 
 exactly the product of the assigned values, i.e., $v_\context= \prod_{k\in 
 \context} a_k$.
The most general noncontextual HV model predicts fixed assignments $\vec 
 a^{(i)}$ with probabilities $p_i$, and hence the set of correlations that can 
 be explained by a noncontextual HV models is characterized by the convex hull 
 of the models with fixed assignments, thus forming the noncontextuality 
 polytope.

Then, a noncontextuality inequality is an affine bound on the noncontextuality 
 polytope, i.e., a real vector $\vec \lambda$ such that $\eta\ge \vec 
 \lambda\cdot \vec v$ for all correlation vectors $v$ that originate from a 
 noncontextual model:
\begin{equation}\label{e:coneinq}
 \eta\ge \sum_{\context\in \Context} \lambda_\context
 \prod_{k\in \context} a_k,
\end{equation}
 for any assignment $\vec a\equiv (a_1, \dotsc, a_n)\in \{-1,1\}^n$.

In quantum mechanics, in contrast, the measurement of the entry $v_\context$ 
 corresponds to the expectation value $\exv{\prod_{k\in \context} A_k}{}_\rho$, 
 where $\rho$ specifies the quantum state.
Thus the value of $\vec \lambda \cdot \vec v$ predicted by quantum mechanics is 
 given by $\exv{T(\vec\lambda)}{}_\rho$, with
\begin{equation}
 T(\vec\lambda)= \sum_{\context\in \Context} \lambda_\context \prod_{k\in 
 \context} A_k.
\end{equation}
If the expectation value exceeds the noncontextual limit $\eta$, then the 
 inequality demonstrates contextual behavior, yielding the quantum violation
\begin{equation}
\mathcal V= \frac {\max_\rho \exv{T(\vec\lambda)}_\rho} \eta-1.
\end{equation}
An inequality is optimal, if the violation is maximal for the given 
 contextuality scenario.
In general, this optimization is difficult to perform and it is not always 
 clear that an optimal inequality also yields the most significant violation 
 \cite{Jungnitsch:2010PRL}.

\emph{Optimal state independent violation and tight inequalities.}---%
However, if we require a state independent violation of the inequality, without 
 loss of generality, $T(\vec\lambda)= \openone$ and hence the optimization over 
 the quantum state $\varrho$ vanishes.
Then, the coefficient vector $\vec\lambda$ and the noncontextuality bound 
 $\eta$ are optimal if $\eta$ is minimal under the constraint $T(\vec\lambda)= 
 \openone$ and if the inequalities in Eq.~\eqref{e:coneinq} are satisfied.
That is, we ask for a solution $(\eta^*, {\vec\lambda}^*)$ of the optimization 
 problem
\begin{equation}\label{e:linpro}\begin{split}
 \text{minimize:}&\quad \eta,\\
 \text{subject to:}&\quad
 T(\vec\lambda)= \openone
 \text{ and}\\&\quad\text{Eq.~\eqref{e:coneinq} holds for all $\vec a$.}
\end{split}\end{equation}
This optimization problem is a linear program and such programs can be solved 
 efficiently by standard numerical techniques and optimality is then 
 guaranteed.
We implemented this optimization using CVXOPT \cite{cvxopt} for Python, which 
 allows us to study inequalities with up to $n=21$ observables and 
 $\abs{\Context}=131$ contexts.
Note that this program also solves the feasibility problem, whether a 
 contextuality scenario exhibits SIC at all.
This is the case, if and only if the program finds a solution with $\eta<1$ and 
 thus $\mathcal V>0$.

The optimal coefficients ${\vec\lambda}^*$ are, in general, not unique but 
 rather form a polytope defined by Eq.~\eqref{e:coneinq} with $\eta=\eta^*$.
This leaves the possibility to find optimal inequalities with further special 
 properties.
There are at least two important properties that one may ask for.
Firstly, from an experimental point of view, it would be desirable to have some 
 of the coefficients $\lambda_\context=0$, since then the context $\context$ 
 does not need to be measured.
In general, it will depend on the experimental setup, which coefficients
 $\lambda_\context=0$ yield the greatest advantage.
For the sequential measurement schemes it is natural to choose the longest 
 measurement sequences.
Secondly, there might be \emph{tight} inequalities among the optimal solutions:
An inequality is tight, if the affine hyperplane given by the solutions of 
 $\eta= \vec \lambda\cdot \vec x$ is tangent to a facet of the noncontextuality 
 polytope.
This property can be readily checked using Pitowsky's construction 
 \cite{Pitowsky:1989}:
Denote by $p$ the affine dimension of the noncontextuality polytope and choose 
 those assignments $\vec a$, for which Eq.~\eqref{e:coneinq} is saturated.
Then, the inequality is tangent to a facet if and only if the affine space 
 spanned by the vertices $\vec v\equiv (\prod_{k\in \context} a_k\mid 
 \context\in \Context)$ is $(p-1)$-dimensional.
 
Furthermore, we mention that the condition of state independence might be 
 loosened to only require that the quantum violation is \emph{at least} 
 $\mathcal V$ for all quantum states.
This corresponds to replacing the condition $T(\vec\lambda)= \openone$ by the 
 condition that $T(\vec\lambda)- \openone$ is positive semidefinite.
Then, the linear program in Eq.~\eqref{e:linpro} becomes a semidefinite 
 program, which still can be solved by standard numerical methods with 
 optimality guaranteed.
However, for the examples that we consider in the following, the semidefinite 
 and the linear program yield the same results.

\begin{figure}%
\centerline{\includegraphics[width=.6\linewidth]{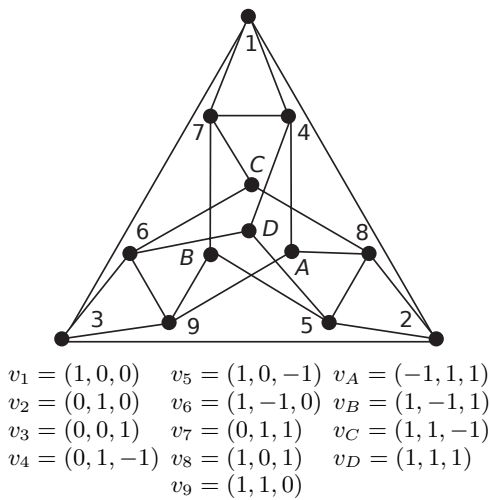}}%
\vspace{.5em}%
\begin{tabular}{lll}
$v_1=(1,0, 0)$&$v_5=(1, 0,-1)$&$v_A=(-1, 1, 1)$\\
$v_2=(0,1, 0)$&$v_6=(1,-1, 0)$&$v_B=( 1,-1, 1)$\\
$v_3=(0,0, 1)$&$v_7=(0, 1, 1)$&$v_C=( 1, 1,-1)$\\
$v_4=(0,1,-1)$&$v_8=(1, 0, 1)$&$v_D=( 1, 1, 1)$\\
              &$v_9=(1, 1, 0)$&
\end{tabular}%
\caption{\label{Fig1}%
Graph of the compatibility relations between the observables for the Yu-Oh
 scenario.
Nodes represent vectors $|v_i\rangle$ [or the observables $A_i$ defined in 
 (\ref{e:obs})] and edges represent orthogonality (or compatibility) 
 relations.}
\end{figure}
\begin{table}
\begin{tabular}{c|c|c|c||c|c|c|c||c|c|c|c}
$\context$&YO&opt$_2$&opt$_3$&
$\context$&YO&opt$_2$&opt$_3$&
$\context$&YO&opt$_2$&opt$_3$\\\hline
$1$&$2$&$2$&$1$&$A\!$--$\!D$&$\phantom-2$&$\phantom-1$&$\phantom-2$&$
3,9$&$-1$&$-2$&$-1$\\
$2$&$2$&$3$&$1$&$1$,$2$&$-1$&$-1$&$-2$&$4,7$&$-1$&$\phantom-0$&$-1$\\
$3$&$2$&$3$&$1$&$1$,$3$&$-1$&$-1$&$-2$&$5,8$&$-1$&$-2$&$-1$\\
$4$&$2$&$1$&$1$&$1$,$4$&$-1$&$-1$&$-1$&$6,9$&$-1$&$-2$&$-1$\\
$5$&$2$&$2$&$1$&$1$,$7$&$-1$&$-1$&$-1$&$*$,$A\!$--$\!D$&$-1$&$-1$&$-2$\\
$6$&$2$&$2$&$1$&$2$,$3$&$-1$&$-2$&$-2$&$1$,$2$,$3$&--&--&$\phantom-0$\\
$7$&$2$&$1$&$1$&$2$,$5$&$-1$&$-2$&$-1$&$1$,$4$,$7$&--&--&$-3$\\
$8$&$2$&$2$&$1$&$2$,$8$&$-1$&$-2$&$-1$&$2$,$5$,$8$&--&--&$-3$\\
$9$&$2$&$2$&$1$&$3$,$6$&$-1$&$-2$&$-1$&$3$,$6$,$9$&--&--&$-3$
\end{tabular}
\caption{\label{tab}%
Coefficients $\lambda_\context$ of inequalities for the Yu-Oh scenario.
The column $\context$ labels the different contexts, YO the coefficients in the 
 inequality of Ref.~\cite{Yu:2012PRL}, opt$_2$ an optimal tight inequality with 
 contexts of maximal size {2}, opt$_3$ an optimal tight inequality with 
 contexts of all sizes.
For compactness, the coefficients in the column YO have been multiplied by 
 $50/3$, for the column opt$_2$ by $52/3$ and for the column opt$_3$ by $83/3$.
The row labeled ``$A$--$D$'' shows the coefficients for the contexts $\{A\}$, 
 $\{B\}$, $\{C\}$, $\{D\}$ and the row labeled ``$*$,$A$--$D$'' shows the 
 coefficients for $\{4,A\}$, $\{8,A\}$, $\{9,A\}$, $\{5,B\}$, $\{7,B\}$, 
 $\{9,B\}$, $\{6,C\}$, $\{7,C\}$, $\{8,C\}$, $\{4,D\}$, $\{5,D\}$, $\{6,D\}$.}
\end{table}
%
\emph{Most fundamental noncontextuality inequalities.}---%
We now apply our method to the SIC scenario for a qutrit system introduced by 
 Yu and Oh \cite{Yu:2012PRL}.
Qutrit systems are of fundamental interest, since no smaller quantum system can 
 exhibit a contextual behavior \cite{Bell:1966RMP, *Kochen:1967JMM}.
It has been shown, that this scenario is the simplest possible SIC scenario for 
 a qutrit \cite{Cabello:2011XXX}.

For a qutrit system, the dichotomic observables are of the form
\begin{equation}\label{e:obs}
 A_i=\openone - 2 \proj{v_i}.
\end{equation}
In the Yu-Oh scenario, there are {13} observables defined by the {13} unit 
 vectors $\ket{v_i}$ provided in Fig.~\ref{Fig1}.
In the according graph, each operator is represented by node $i\in V$ of the 
 graph $G=(V,E)$ and an edge $(i,j)\in E$ indicates that $\ket{v_i}$ and 
 $\ket{v_j}$ are orthogonal, $\bracket{v_j}{v_i}=0$, so that $A_i$ and $A_j$ 
 are compatible.
The original inequality takes into account all contexts of size one and two, 
 $\Context_\mathrm{YO}=\{\{1\}, \dotsc, \{D\}\}\cup E$ and the coefficients 
 were chosen to $\lambda_\context=-3/50$ if $\context \in E$ and 
 $\lambda_\context=6/50$ else.
This yields an inequality with a state-independent quantum violation of 
 $\mathcal V= 1/24\approx 4.2\%$.

With the linear program we find that the maximal violation for the contexts 
 $\Context_\mathrm{YO}$ is $\mathcal V= 1/12\approx 8.3\%$ and thus twice that 
 of the inequality in Ref~\cite{Yu:2012PRL}.
Interestingly, among the optimal coefficients $\vec \lambda^*$ there is a 
 solution which is tight and for which the coefficients $\lambda_{4,7}$ 
 vanishes, cf.\ Table~\ref{tab}, column ``opt$_2$'' for the list coefficients.
We find that up to symmetries, $\lambda_{4,7}$ is the only context that can be 
 omitted while still preserving optimality.

In order to demonstrate the practical advantage, let us discuss the recent 
 experimental values obtained for the Yu-Oh scenario \cite[FIG.~2]{Zu:2012PRL}.
For those values, the original Yu-Oh inequality is violated by {3.7} standard 
 deviations.
But if the same data is evaluated using our optimal inequality ``opt$_2$'', the 
 violation increases to {7.5} standard deviations.
We mention, however, that the particular experimental setup implements the same 
 observable in different context differently, thus easily allowing a 
 noncontextual HV model explaining the data \cite{Guhne:2010PRA}.
A setup avoiding such problems is described in Ref.~\cite{Cabello:2012PRA}.

The maximal contexts in the Yu-Oh scenario are of size three, and hence it is 
 possible to include also the corresponding terms in the inequality, i.e., we 
 extend the contexts $\Context_\mathrm{YO}$ by the contexts $\{1,2,3\}$, 
 $\{1,4,7\}$, $\{2,5,8\}$, and $\{3,6,9\}$.
Since this increases the number of parameters in the inequality, there is a 
 chance that this case allows an even higher violation.
In fact, the maximal violation is $\mathcal V= 8/75\approx 10.7\%$.
Again, it is possible to find tight inequalities with vanishing coefficients, 
 and in particular the context $\{1,2,3\}$ can be omitted; the list of 
 coefficients is given in Table~\ref{tab}, column ``opt$_3$''.

\emph{Further examples.}---%
Our method is applicable to all SIC scenarios, providing the optimal 
 inequality.
We mention two further examples:
(i) The ``extended Peres-Mermin square'' uses as observables all 15 products of 
 Pauli operators on a two-qubit system, $(\sigma_\mu\otimes \sigma_\nu)$ 
 \cite{Cabello:2010PRA}.
The optimal violation is $\mathcal V = 2/3$, where only contexts of size three 
 need to be measured and $\lambda_\context=1/15$, except 
 $\lambda_{xx,yy,zz}=\lambda_{xz,yx,zy}=\lambda_{xy,yz,zx}=-1/15$.
Among the optimal solutions no simpler inequality exists.
 (ii) The {18} vector proof \cite{Cabello:1996PLA} of the Kochen-Specker 
 theorem uses a ququart system and {18} observables of the form \eqref{e:obs}.
For contexts up to size {2} the maximal violation is $\mathcal V= 1/17\approx 
 5.9\%$ (cf.\ \cite{Yu:2011XXX}), while including all context the maximal 
 violation is $\mathcal V= 2/7\approx 28.6\%$ (cf.\ \cite{Cabello:2008PRL}).
The situation where only contexts up to size {3} are admissible has not yet 
 been studied and we find numerically a maximal violation of $\mathcal V 
 \approx 14.3\%$.

\emph{Conclusions.}---%
Contextuality is suspected to be one of the fundamental phenomena in quantum 
 mechanics.
While it can be seen as the underlying property of the nonlocal behavior of 
 quantum mechanics, so far no methods for a systematic investigation have been 
 developed.
We here showed that Pitowsky's polytope naturally generalizes to the 
 noncontextual scenario and hence the question of a full characterization of 
 this noncontextuality polytope arises.
This can be done via the so-called tight inequalities.
On the other hand, among the most striking aspects where contextuality is more 
 general than nonlocality is that the former can be found to be independent of 
 the quantum state.
For this state-independent scenario, we showed that the search for the optimal 
 inequality reduces to a linear program, which can be solved numerically with 
 optimality guaranteed.
We studied several cases of this optimization and find that in all those 
 instances one can construct noncontextuality inequalities with a state 
 independent violation that are, in addition, tight.
This is in particular the case for the most fundamental scenario of state 
 independent contextuality \cite{Yu:2012PRL} and we presented two essentially 
 different inequalities---one involves at most contexts of size two, the other 
 of size three.
We hence lifted the Yu-Oh scenario to the same fundamental status as the CHSH 
 Bell inequality \cite{Clauser:1969PRL, *Clauser:1970PRL}, which is the 
 simplest scenario for nonlocality.
Our state-independent tight inequalities are particularly suitable for 
 experimental tests and hence we expect that they stimulate experiments to 
 finally observe SIC in qutrits \cite{Cabello:2012PRA}.

\begin{acknowledgments}
The authors thank J.~R.\ Portillo for checking some calculations.
This work was supported by the Spanish Project No.\ FIS2011-29400, the EU 
 (Marie-Curie CIG 293933/ENFOQI), the Austrian Science Fund (FWF): Y376-N16 
 (START prize), and the BMBF (CHIST-ERA network QUASAR).
\end{acknowledgments}

\bibliography{the,local}

\end{document}